\documentclass[aps,prl,twocolumn,showpacs,amsmath,floatfix]{revtex4}
\usepackage{amsmath,graphics,epsfig,color,verbatim,ulem}

\begin{document}

\title{Prediction of switchable half semiconductor in d$^{1}$ transition metal dichalcogenide monolayers}

\author{Peng-Ru Huang}
\affiliation{Department of Physics, Yunnan University, Kunming 650091, China}
\author{Yao He}
\email{yhe@ynu.edu.cn}
\affiliation{Department of Physics, Yunnan University, Kunming 650091, China}
\author{Hridis K. Pal}
\affiliation{School of Physics, Georgia Institute of Technology, Atlanta, Georgia 30332, USA}
\author{Markus Kindermann}
\affiliation{School of Physics, Georgia Institute of Technology, Atlanta, Georgia 30332, USA}

\marginparwidth 2.7in
\marginparsep 0.5in

\begin{abstract}

We propose that a half semiconducting state can exist in trigonal-prismatic transition metal dichalcogenide (TMDC) monolayers of d$^{1}$ configuration. In that state both electrons and holes are spin polarized and share the same spin channel. On the basis of hybrid density functional theory, we predict in particular that VS$_2$ monolayers are half semiconductors with a direct band gap. Moreover, we find that the conduction electron spin orientation of VS$_2$ switches under moderate strain. Our predictions thus open up intriguing possibilities for applications of VS$_2$ in spintronics and optoelectronics. Our analysis of trigonal-prismatic group-V MX$_2$ (M=V, Nb, Ta; X=S, Se, Te) monolayers reveals a broad diversity of electronic states that can be understood qualitatively in terms of localization of $d$ electrons.

\end{abstract}
\pacs{73.20.At,73.22.-f,75.30.-m}

\maketitle

Two-dimensional transition metal dichalcogenides (TMDCs) have been the subject of intense research over the past decade because of their unusual electronic properties and potential applications in future nanoelectronic devices \cite{MS1,MS2,MS3}. The prototype of such materials is MoS$_2$, which undergoes a transition from an indirect band gap in the bulk to a direct band gap in the monolayer \cite{MS4}. The direct-band-gap transition, together with its valley and spin selective properties, makes the material very promising for new generation electronics \cite{MS5,MS6}. Related properties were also observed in other group-VI dichalcogenides such as MoSe$_2$, WS$_2$ and WSe$_2$ \cite{VI1,VI2,VI3}. In comparison, the central topic in group-V TMDCs is electronic instabilities \cite{V1,V2,V3}. Owing to complicated electron-electron and electron-phonon interactions, this series of compounds displays a rich phase diagram which includes metal, charge density wave (CDW), Mott insulator, and even superconductor \cite{V4}. In addition, their electronic structure appears to be highly sensitive to temperature and pressure, which can induce CDW or metal-insulator transitions \cite{V5}. The complexity of the electronic structure of these materials holds promise of more surprising physics yet to be discovered.

\begin{figure}
\includegraphics[clip,width=3.0 in]{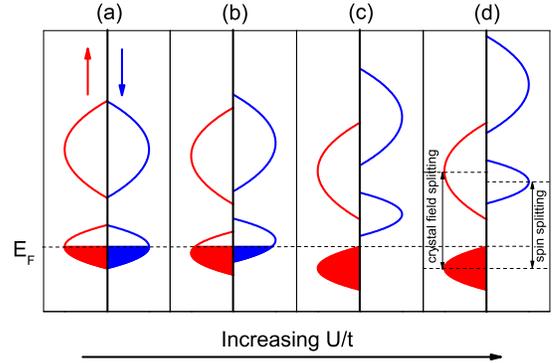}
\caption{\label{Fig:1}(Color online) Energy band diagrams for trigonal-prismatic group-V MX$_2$ monolayers with increasing spin splitting of valence band. Depending on $U/t$, four different mean-field phases are expected: (a) nonmagnetic metal, (b) magnetic metal, (c) magnetic semiconductor, and (d) half semiconductor.}
\end{figure}

In this Letter, we study the electronic structures of trigonal-prismatic MX$_2$ (M=V, Nb, Ta; X=S, Se, Te) monolayers based on first-principle calculations at the level of the hybrid functional, and propose that the ground state of VS$_2$ monolayers is a correlation-driven half semiconducting state. In this intriguing half semiconducting state --- examples of which are rare --- both the valence band maximum (VBM) and the conduction band minimum (CBM) are spin-polarized, and their spin orientations are identical. Moreover, the spin sign of the CBM in VS$_2$ is found to switch under moderate strain, undergoing a phase transition from a half semiconductor to a magnetic semiconductor. This opens up intriguing possibilities for applications in novel spin based electronics. Our analysis of other Group-V MX$_2$ monolayers (M=V, Nb, Ta; X=S, Se, Te) reveals a broad range of correlation-driven phases such as ferromagnetic metallic, ferromagnetic semiconducting states, and half semiconducting states (Fig. \ref{Fig:1}). We demonstrate that this rich phase diagram --- interesting both from a fundamental as well as an applied perspective --- has an intuitive interpretation in terms of localization of $d$ electrons.

\begin{figure}
\includegraphics[clip,width=3.0 in]{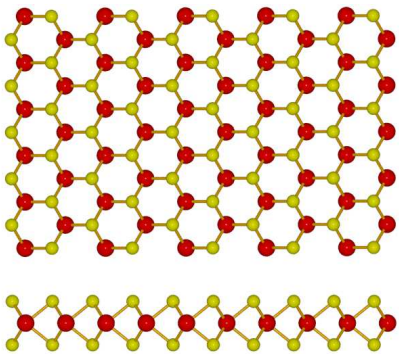}
\caption{\label{Fig:2}(Color online) Top view and side view of trigonal-prismatic MX$_{2}$ monolayers. (M atoms in red and X atoms in yellow).}
\end{figure}

A trigonal-prismatic MX$_2$ monolayer is composed of one triangular transition-metal sublattice sandwiched between two sheets of chalcogen atoms. Chalcogen atoms are below and above the centers of metal triangles, resulting in a six-fold trigonal-prismatic coordination for the metal , cf. Fig. \ref{Fig:2}. The coordination splits the metal $d$ orbitals into $A^{\prime}_{1}$ ($d_{z^{2}}$), $E^{\prime}$ ($d_{xy}$,$d_{x^{2}-y^{2}}$), and $E^{\prime\prime}$ ($d_{xz}$,$d_{yz}$) manifolds \cite{E1}. Governed by the overlap with chalcogen orbitals, their mean energies lie in ascending order of $A^{\prime}_{1}$, $E^{\prime}$, and $E^{\prime\prime}$. The $d_{z^{2}}$, $d_{xy}$, and $d_{x^{2}-y^{2}}$ orbitals further hybridize with each other owing to common reflection symmetry, contributing a narrow valence subband and a wider conduction subband \cite{E2,E3}. In other words, the low energy bands of MX$_2$ are dominated by the onsite hybrid and the inter-site overlap of the $d_{z^{2}}$, $d_{xy}$, and $d_{x^{2}-y^{2}}$ orbitals of the 2D metal sublattice. A tight-binding model taking into account these three orbitals already captures well the dispersion of the low energy bands --- especially the valence band \cite{TB}. Also recent experimental investigations provide evidence for the metal-metal interactions to define the electronic structure of such 2D systems \cite{V3}. For these reasons, a multi-orbital Hubbard model on the triangular lattice is a good candidate for describing the strongly correlated physics of d$^{1}$ MX$_2$ monolayers. However, to gain a qualitative understanding of the emergence of the various correlation-driven phases shown in the schematic diagrams of Fig. \ref{Fig:1}, we use a simpler, single band Hubbard model:

\begin{equation}\label{Eq:1}
H=-t\sum_{i,j,\sigma}{(c^{+}_{i\sigma}c_{j\sigma}+c^{+}_{j\sigma}c_{i\sigma})}+U\sum_{i}{n_{i\uparrow}n_{i\downarrow}} .
\end{equation}	

Here, $t$ is the nearest-neighbor hopping integral, and $U$ is the on-site coulomb potential. $c^{+}_{i\sigma}$($c_{i\sigma}$) creates (annihilates) an electron of spin  $\sigma$ at site $i$, and  $n_{i\sigma}$ is the number operator for spin  $\sigma$ at site $i$. $U$ and $t$ are two competing parameters that parametrize the localization and the mobility of the strongly correlated electrons, respectively. The onsite repulsion $U$ is responsible for the spin splitting of the narrow $d$ band. As illustrated in Fig.\ref{Fig:1}, on a mean-field level, there are four possible band structure configurations, depending on the magnitude of the spin splitting. Without any spin splitting (Fig. \ref{Fig:1} a) the system is a half-filled nonmagnetic metal. In the second case (Fig. \ref{Fig:1}b) the two spin components of the narrow $d$ band are partially separated and the system is a magnetic metal. As the spin splitting starts exceeding the bandwidth, such that the two spin components are fully separated (Fig. \ref{Fig:1}c), a magnetic semiconductor results. Finally, when the spin splitting of the narrow $d$ band becomes larger even than the crystal field splitting, so that one spin component of the narrow $d$ band is above the CBM, which is typically about 1.0 eV higher than the VBM for TMDCs, the system is a half semiconductor (Fig. 1d): both the VBM and the CBM are spin-polarized and have the same spin orientation. The first three states are frequently reported while the fourth one is rare in nature. Interestingly, as our calculations show, d$^{1}$ TMDCs span the entire gamut of phases shown in Fig. \ref{Fig:1}, including the half semiconducting state, which we predict to exist in VS$_2$. We note that a large spin splitting comparable to the crystal field splitting does not always result in a half semiconductor: the VBM and CBM originating from two different spin-split bands may end up having different spins (as in the case of VSe$_2$ and VTe$_2$; see below).

\begin{table}%[b]
\caption{Electronic states for trigonal-prismatic MX$_2$ (M=V, Nb, Ta; X=S, Se,Te) computed at the level of hybrid functional DFT. *Although VSe$_{2}$ and VTe$_{2}$ are magnetic semiconductors, their VBM and CBM arise from splitting of different bands, similar to half semiconductors, unlike Fig. \ref{Fig:1}c. }
\begin{ruledtabular}
\begin{tabular}{c|c c c}
Composition  & V          & Nb   & Ta   \\
\hline
S            & HS         & MM   & MM   \\
Se           & MS$^{*}$   & MS   & MS   \\
Te           & MS$^{*}$   & MS   & MS   \\
\end{tabular}
\end{ruledtabular}
\label{Tab:1}
\end{table}

We make the above predictions for the electronic structures of trigonal-prismatic MX$_2$ (M=V, Nb, Ta; X=S, Se, Te) monolayers based on first-principles calculations. (We note that although for VSe$_2$ and VTe$_2$ the trigonal-prismatic configuration is not the ground state, we investigate it here for its theoretical interest.) The calculations are performed using density functional theory as implemented in the Vienna ab initio simulation package (VASP) \cite{vasp}. The exchange-correlation functional is considered at the level of the generalized gradient approximation (GGA) \cite{PBE}. The screened hybrid functional method proposed by Heyd, Scuseria, and Ernzerhof (HSE) \cite{hse} is used. The single-particle equations are solved using the projector-augmented wave (PAW) method \cite{paw1,paw2} with a plane-wave basis and a cutoff energy of 500 eV. All the calculations are carried out in a slap model with the primary surface cell without considering any CDW distortion, since we focus on the electron-electron interaction only. Our results are summarized in Table \ref{Tab:1}. While all three VX$_{2}$ (X=S, Se, and Te) materials show large spin splitting comparable to the crystal field splitting, only VS$_2$ has the same spin orientation for both CBM and VBM, and, therefore, is a half semiconductor. In trigonal-prismatic VSe$_{2}$ and VTe$_{2}$ they have different spin orientations (see Supplementary materials). Like NbSe$_{2}$, NbTe$_{2}$, TaSe$_{2}$, and TaTe$_{2}$ they are thus ferromagnetic semiconductors, while NbS$_{2}$ and TaS$_{2}$ are ferromagnetic metals.

\begin{figure}
\includegraphics[clip,width=3.0 in]{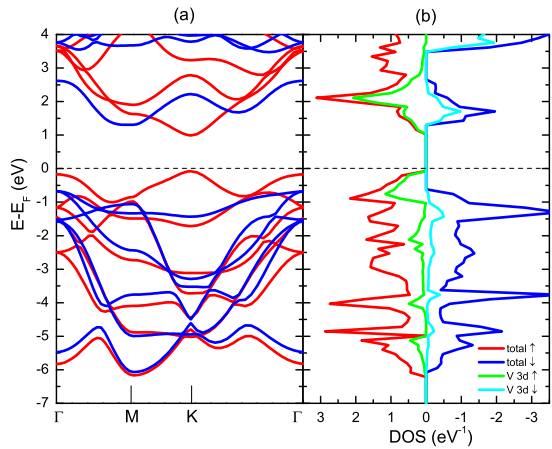}
\caption{\label{Fig:3}(Color online) Spin-resolved (a) band structures and (b) density of states for VS$_2$ computed with hybrid DFT. The presented V 3$d$ density of states are contributed by $d_{z^{2}}$, $d_{xy}$, and $d_{x^{2}-y^{2}}$ orbitals.}
\end{figure}

{\it Half semiconducting state and direct band gap in VS$_2$}. According to our total energy calculations, the VS$_2$ monolayer is in a ferromagnetic ground state, consistent with both prior calculations \cite{vx1} and experimental observations \cite{vx2,vx3}. The ferromagnetic state is more stable than both the nonmagnetic and the antiferromagnetic states (spins of nearest neighbors are opposite), by 367 meV and 689 meV, per cell, respectively. The magnetic moment per unit cell is calculated to be 1.0 $\mu B$. Each V atom contributes 1.30 $\mu B$, while each S atom contributes -0.15 $\mu B$. As is seen from Fig. \ref{Fig:3}, the electrons near both the VBM and the CBM have the same spin orientation, making VS$_2$ monolayers moreover half semiconductors. The highest valence subband is made up of spin-up hybrid states of $d_{z^{2}}$, $d_{xy}$, and $d_{x^{2}-y^{2}}$. The Coulomb interaction causes a $\sim$2.5 eV spin splitting, pushing the down-spin counter part of this band into the conduction band. The CBM consists of spin-up electronic states of the higher hybrid band. As a result, the total density of states matches well the half semiconducting state illustrated by the schematic Fig. \ref{Fig:1}d. In addition, both the VBM and CBM are located at the $K$ point, with a direct band gap of $\sim$1.1 eV. The highest occupied state at $K$ is 93 meV higher than the one at the $\Gamma$ point. The lowest unoccupied state at $K$ is 316 meV lower than that of the opposite spin at a local minimum near $M$. The direct band gap of the VS$_2$ monolayer is similar in magnitude to that of the isostructure MoS$_2$ monolayer, and it is well in the region of the visible spectrum --- of great interest for optoelectronic applications. We arrive at similar conclusions by performing additional, GGA+U calculations (see Supplementary materials).

Experimentally, VS$_2$ monolayers have not yet been realized. So far only thin films of VS$_2$ have been reported \cite{vs1}. They are metallic, as a result of being in the octahedral phase and because of the interlayer interaction \cite{tx}. Further experimental efforts are needed to confirm the intriguing properties we predict for VS$_2$ monolayers.

\begin{figure}
\includegraphics[clip,width=3.0 in]{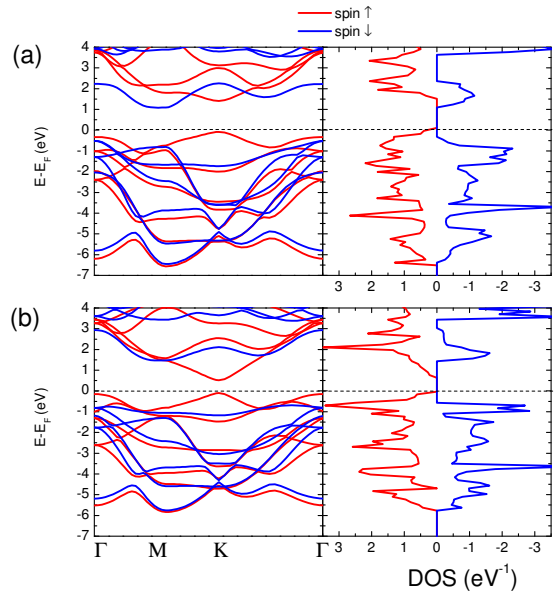}
\caption{\label{Fig:4}(Color online) Spin-resolved electronic structures for (a) 3\% contracted and (b) 3\% expanded VS$_2$.}
\end{figure}

{\it Strain induced phase transition in VS$_2$ monolayer}. Mechanical strain provides a means to tune the electronic structure of 2D electronic systems by controlling the spacing between atoms. We, therefore, next examine the effect of strain on the electronic structure of VS$_2$. According to our calculations, compressive strain tends to decrease the energy of the half semiconducting state and destroy it while tensile strain tends to stabilize it. Fig. \ref{Fig:4} shows the band structure and density of states of a VS$_2$ monolayer under 3\% strain, i.e. the lattice constant was contracted or expanded by 3\%. Under compressive strain, the atomic spacing of V atoms decreases. This increases the orbital overlap and hence the hopping integrals $t$ of the $d$ electrons. As a result, the width of the narrow $d$ band is increased owing to the electron delocalization and the spin splitting is reduced. The system transitions into the normal ferromagnetic semiconducting state due to the lowering of the energy of the down spin counterpart of the valence $d$ band (Fig. \ref{Fig:4}a). The lowest unoccupied state then is a down spin state near $M$, which has energy 327 meV lower than the lowest state at $K$, which is spin-up. According to our calculations, the critical strain for this transition is 2\%. This is a quite moderate, experimentally achievable value. On the other hand, tensile strain enlarges the atomic spacing and enhances the localization of the metal $d$ electrons at the lattice sites. That localization in turn leads to a larger onsite interaction, which enlarges the energy splitting between opposite spins. Consequently, the half semiconducting state is enhanced as the down spin counterpart of the valence subband is lifted deeper into the conduction band (Fig. \ref{Fig:4}b). Our results reveal that the spin splitting of the $d$ electronic states in VS$_2$ is sensitive to the atomic spacing of V atoms owing to electron localization. Thus, the spin sign of the conduction band can be switched by moderate, compressive strain. This phenomenon has direct application in spin-based electronics.

{\it Electronic phases of group-V MX$_2$ monolayers}. While VSe$_{2}$ and VTe$_{2}$ have similar spin splitting as VS$_2$, pushing the down spin counterpart of the $d$ valence subband into the conduction band, the VBM in those cases is not located in the spin-up $d$ subband, but in a spin-split subband arising from deeper $p$ bands. The VBM and CBM thus have different spin signs (see Supplementary materials), such that these materials are unconventional ferromagnetic semiconductors. On the other hand, NbS$_{2}$ and TaS$_{2}$ are ferromagnetic metals, and NbSe$_{2}$, NbTe$_{2}$, TaSe$_{2}$, and TaTe$_{2}$ are normal ferromagnetic semiconductors. This series spans the entire range of strongly correlated phases illustrated in Fig.\ref{Fig:1}. A summary of the phases is given in Table \ref{Tab:1}.

\begin{figure}
\includegraphics[clip,width=3.0 in]{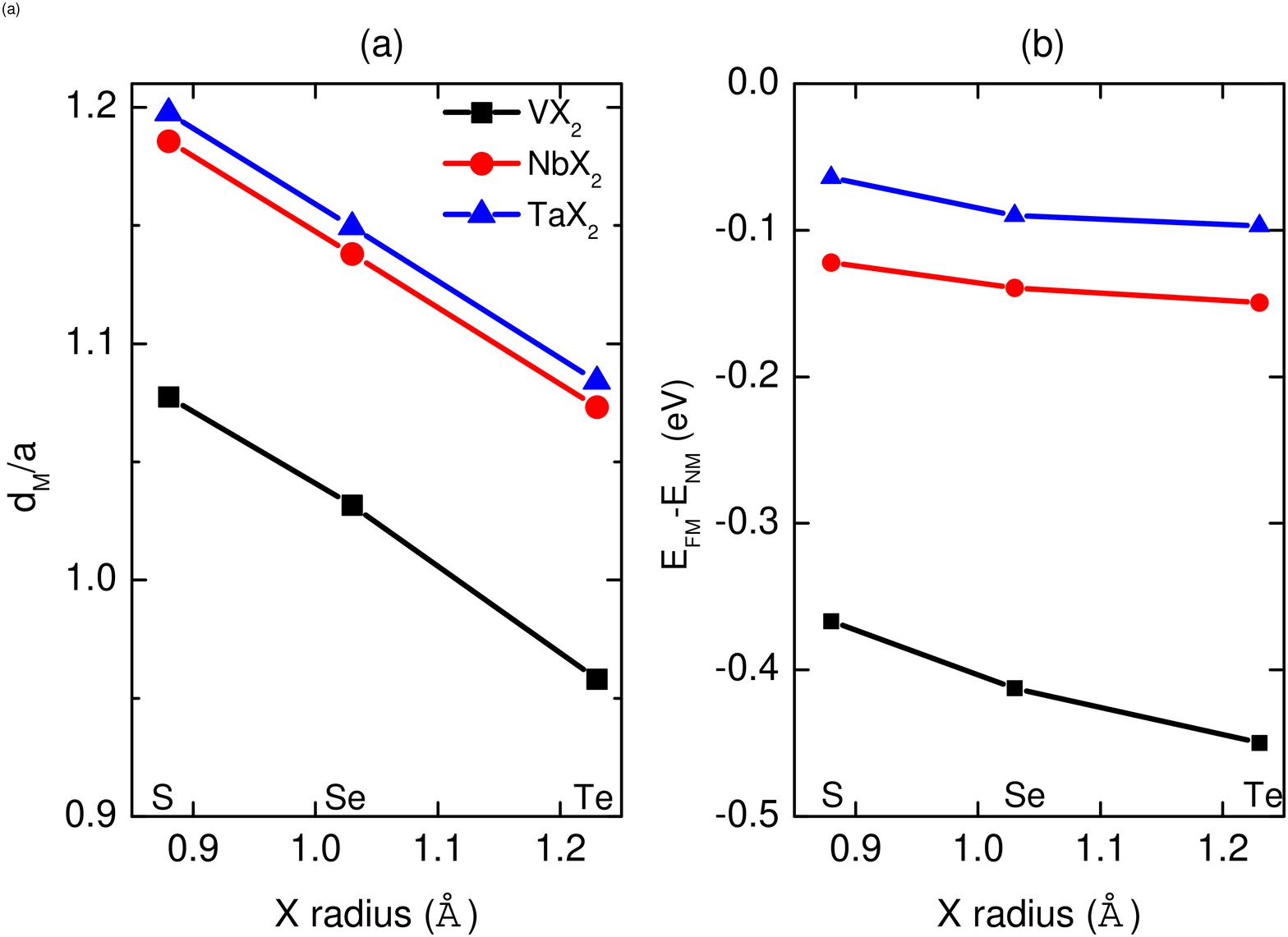}
\caption{\label{Fig:5}(Color online) (a) The ratio of M diameter to the cell constant, $d_{M}/a$, and (b) the difference between the energies $E_{FM}$ of the ferromagnetic and $E_{NM}$ of the nonmagnetic states as a function of X radius for MX$_{2}$. (M=V, Nb, Ta; X=S, Se, Te.).}
\end{figure}

There is a trend for small metal and large chalcogen components to stabilize the phase on the large $U/t$ side of our Hubbard mean-field schematic Fig. \ref{Fig:1}, which can be attributed to electron localization. Consider MX$_2$ of the same metal but different chalcogen components X. In this case, the orbital properties of the metal components are fixed parameters, while the hopping integral is determined by the distance between the transition metal atoms. As shown in Fig. \ref{Fig:5}a, the ratio of metal diameter to the cell constant, $d_{M}/a$, decreases with increasing chalcogen radius from S to Te. Since this ratio qualitatively measures the overlap of the $d$ orbitals, this results in a narrowing of the $d$ band and an increase of the onsite interaction, i.e. a larger spin splitting. The phase variation and the trend of the energy gain by ferromagnetic order,  $\Delta E=E_{FM}-E_{NM}$ (Fig. \ref{Fig:5}b) for MX$_2$ with different chalcogen components can be understood by these simple considerations. If one considers instead different metal components keeping the chalcogen component fixed, one can draw a similar conclusion (see Supplementary materials).

The above results provide strong evidence for the strongly correlated nature of electronic states in group-V TMDCs. So it is not surprising that contradicting predictions exist in the literature. Based on the GGA method, Pan recently reported that VX$_2$ monolayers are all semimetallic and ferromagnetic \cite{vx4}. Wasey et al. predict VS$_2$ to be a semimetal, and VSe$_2$ and VTe$_2$ to be semiconductors with indirect band gaps \cite{vx5}. Li et al. also predicted a semiconducting state for VSe$_2$ \cite{vx6}. These variations in the results can be addressed by asking the following question: are the opposite spin components of the narrow $d$ band in contact or fully separated? It is hard to give a conclusive answer at the level of the GGA method since the calculated band structure is close to the transition between the two. Our calculations based on the hybrid functional clearly demonstrate the existence of the half semiconducting state.

It should be noted that the predicted correlation driven ferromagnetic and insulating states are hard to observe in experiments as they are easily corrupted by doping or disorder \cite{d1,d2,d3}. The ferromagnetic, semiconducting state is based on a correlation-driven electronic configuration where the down spin component of the highest valence subband is a few eV higher than its up-spin counterpart and unoccupied. The energy gain over the nonmagnetic and metallic states from the spin exchange, however, is only a hundred meV or smaller. Occupation of the down-spin band would thus be energetically unfavorable, and doping collapses the system into the nonmagnetic and metallic state. Especially for NbX$_2$ and TaX$_2$ the stabilization energy for ferromagnetic order and the semiconducting state is on the order of only tens of meV, making these states likely too fragile to be observed. However, it is possible for the ferromagnetic and semiconducting states to survive in VX$_2$ since they have relatively large stabilization energies on the order of a hundred meV. This may explain why ferromagnetism has been observed in VS$_2$ multilayers.

In summary, we predict that the electronic states of d$^{1}$ TMDC monolayers span a rich variety of correlation-driven phases including magnetic metal, magnetic semiconductor, and half semiconductor. Our predictions are based on hybrid functional DFT calculations. Most importantly, our calculations demonstrate that the VS$_2$ monolayer is a direct band gap half semiconductor, the electron spin orientation of which can be switched by moderate strain. Such easily tunable phase transitions are highly desirable and hold great promise for applications in spintronics. Our study moreover reveals the importance of the role of electron localization in the electronic structure of d$^{1}$ TMDCs, which is of fundamental interest in the theoretical understanding of these materials. We anticipate our findings to inspire further experimental and theoretical work, particularly with a view to application in novel electronics.

This work was supported by National Natural Science Foundation of China (Grant No. 61366007 and No. 11164032 and No. 61066005), Program for New Century Excellent Talents in University (Grant No. NCET-12-1080), Applied Basic Research Foundation of Yunnan Province (Grant No. 2011CI003 and No. 2013FB007), Program for Excellent Young Talents in Yunnan University and by the NSF under grant DMR-1055799 (M. K. and H. K. P.).

{\it Note added}: Shortly before completion of this manuscript we became aware of complementary work on VS$_2$, which confirms that the trigonal prismatic phase of the material studied here is indeed the ground state at the magnitude of strain considered here \cite{vshse}.

\section{supplementary materials}

\begin{figure}
\includegraphics[clip,width=3.0 in]{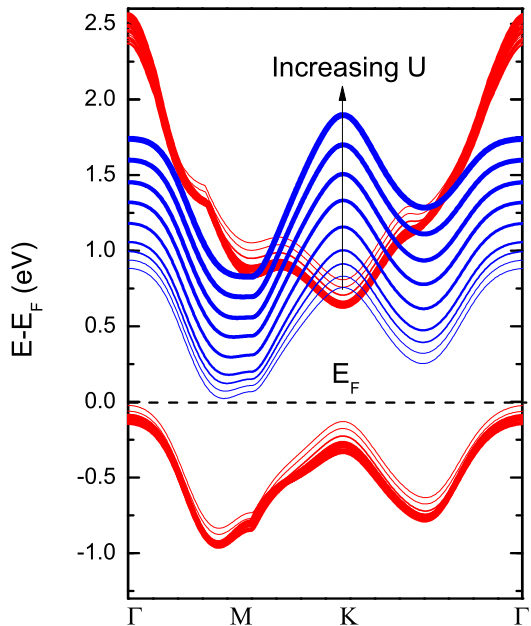}
\caption{\label{Fig:S1}The low energy band structure of VS$_{2}$ as a function of U between 0 and 4.0 eV (0.5 eV per step). The thickness of the line indicates the value of U.}
\end{figure}

{\it GGA+U calculations for VS$_{2}$}. In addition to hybrid calculations, we also carried out GGA+U calculations to examine the effect of Hubbard $U$ on the electronic structure of the VS$_{2}$ monolayer. According to our band structure calculations, the spin splitting of the narrow $d$ band gradually when the onsite Coulomb potential $U$ is increased from 0 to 4.0 eV (Fig. \ref{Fig:S1}). The energy of the down spin component of the valence subband continually increases from that of its spin up counterpart until it is well above the CBM. As a result, the system undergoes a series of transitions, from semimetal to ferromagnetic semiconductor to half semiconductor, as proposed in Fig. \ref{Fig:1}. For a certain value of $U$ ($\sim$3.5 eV), the GGA+U method results in the same prediction of a half semiconducting state as made by the HSE functional. The HSE functional includes some short range Hartree-Fock exchange, which partially cancels the self-interaction emerging in GGA and results in an increased localization of $d$ electrons. The localization of electrons on atomic sites thus increases the onsite interaction. On the other hand, the GGA+U method imposes an extra onsite interaction, leading to larger spin splitting. The consistency of the two approaches demonstrates the critical role of correlation due to localization in the electronic structure of MX$_{2}$.

\begin{figure}
\includegraphics[clip,width=3.0 in]{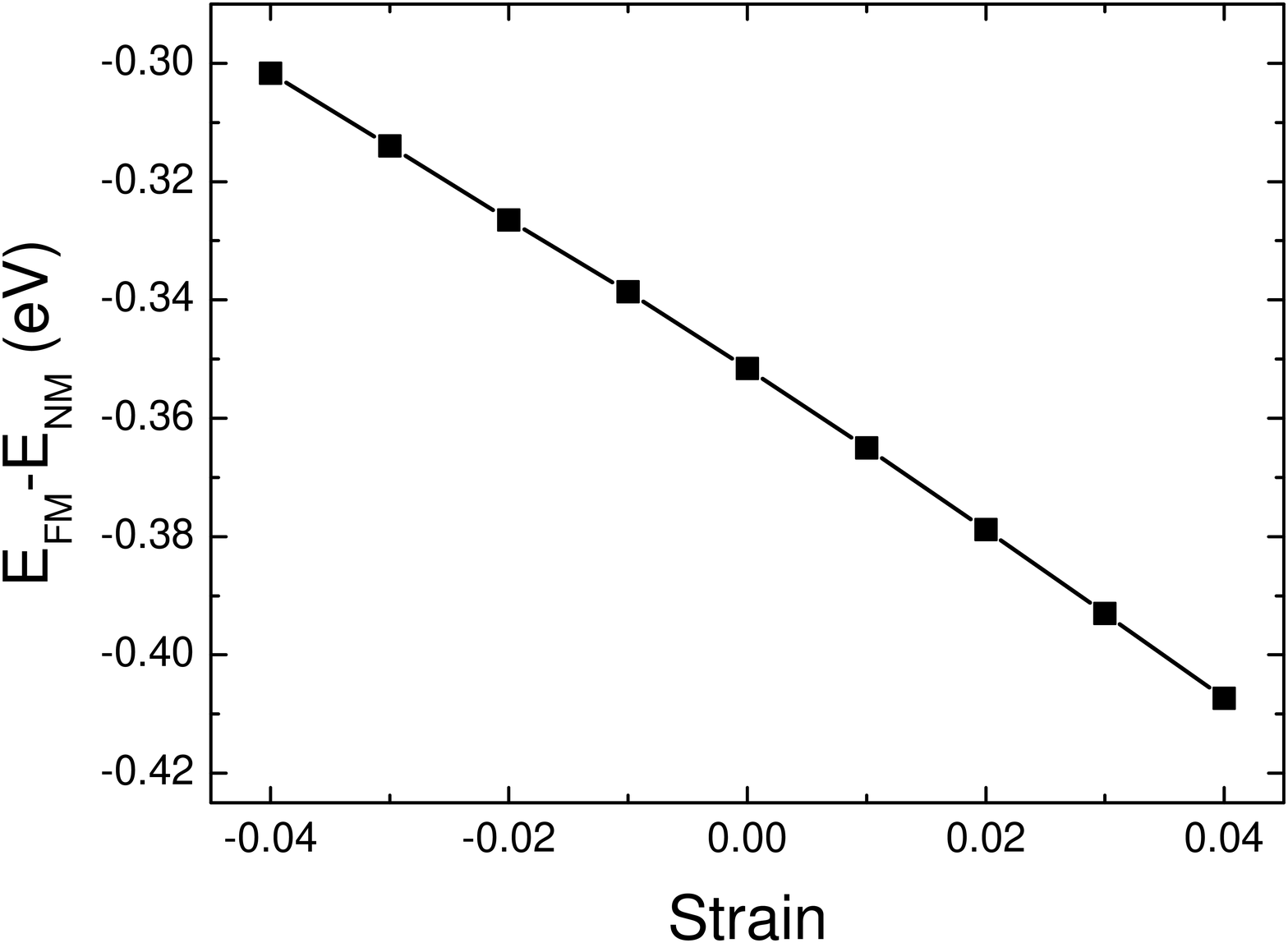}
\caption{\label{Fig:S2}Energy difference between the ferromagnetic and nonmagnetic states of monolayer VS$_{2}$ as a function of strain.}
\end{figure}

{\it Strain and ferromagnetic ordering}. In Fig. \ref{Fig:S2}, we show the energy gain of ferromagnetic order versus the nonmagnetic states $E_{FM} - E_{NM}$ as a function of in-plane strain for monolayer VS$_{2}$. As the strain varies from -4\% (compressive) to 4\% (tensile), the energy gain increases almost linearly from 0.30 eV to 0.41 eV. This demonstrates explicitly that compressive strain decreases the energy gain of ferromagnetic order and tends to destroy the half semiconducting state. On the other hand, tensile strain tends to enhance ferromagnetic ordering and hence the half semiconducting state.

\begin{figure}
\includegraphics[clip,width=3.0 in]{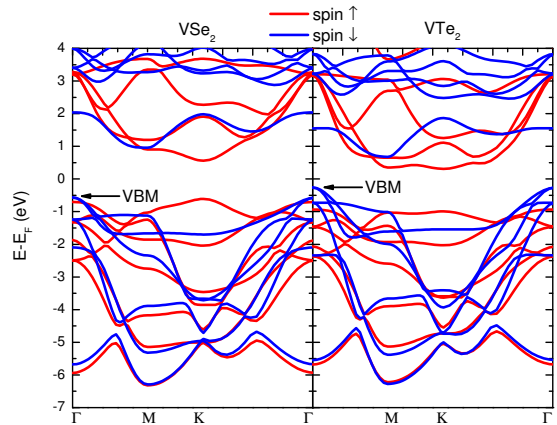}
\caption{\label{Fig:S3}Spin-resolved band structures for monolayer VSe$_{2}$ and VTe$_{2}$.}
\end{figure}

{\it Unconventional magnetic semiconductors, VSe$_{2}$ and VTe$_{2}$}. In Fig. \ref{Fig:S3}, we show the spin-resolved band structures for trigonal-prismatic , monolayer VSe$_{2}$ and VTe$_{2}$, respectively. The down spin counter part of the valence $d$ band is above the CBM in these two compounds. However, in both cases, the VBM is not located in the up spin $d$ band, but in a spin split subband arising from deeper $p$ bands. Therefore, VSe$_{2}$ and VTe$_{2}$ are unconventional magnetic semiconductors which differ from both half semiconductors like VS$_{2}$ and magnetic semiconductors like NbSe$_{2}$, NbTe$_{2}$, TaSe$_{2}$, and TaTe$_{2}$. For VSe$_{2}$, the VBM, made up of $p$ bands, is only 40 meV higher than the highest $d$ states at $K$. It may be possible to tune it into a half semiconducting state. However, for VTe$_{2}$, the VBM is 731 meV higher, deviating greatly from a half semiconducting state.

{\it Effect of metal atoms on localization and ferromagnetic ordering}. The phases in Table \ref{Tab:1} can be understood in terms of electronic localization on the metal atoms. In the main text we have shown how the different phases for the same metal atom with different chalcogen atoms can be understood within this picture. Here we consider the phase differences among MX$_{2}$ materials with the same chalcogen atom, but different metal atoms.  The diameters of the group-V transition elements, V (3d), Nb (4d), and Ta (5d), are 3.42, 3.96, and 4.00 {\AA}, respectively. According to our calculation, the cell constants for their compounds with sodium, for example, are 3.17, 3.35, and 3.35 {\AA} respectively. As a result, the $d_{M}/a$ ratio for VS$_{2}$ (1.08) is much smaller than for NbS$_{2}$ (1.19) and TaS$_{2}$ (1.20). Similar considerations apply for monolayers with Se or Te components. These comparisons show that the $d$ electrons in VX$_{2}$ are more localized than those in NbX$_{2}$ and TaX$_{2}$. Consequently, all VX$_{2}$ monolayers are well stabilized in the semiconducting state, while NbX$_{2}$ and TaX$_{2}$ fall into the critical region of metal to insulator transition. Also, the stabilization energy of ferromagnetic order for VS$_{2}$, $\Delta E$, is 367 meV, much larger than those of NbS$_{2}$ (122 meV) and TaS$_{2}$ (64 meV), and it follows the same trend: the spin splitting in VS$_{2}$ is larger than that in NbS$_{2}$ and TaS$_{2}$ owing to stronger electronic localization.

\end{document}